\documentclass[aps,twocolumn,groupedaddress,showpacs]{revtex4}
\usepackage[dvips]{graphicx}
\usepackage{amsmath}
\usepackage{amssymb}
\begin{document}
\bibliographystyle{prsty}
\title{A Note on Perturbations During a Regular Bounce}
\author{Thorsten~J. Battefeld $^{1)}$\footnote{battefeld@physics.brown.edu}
and  Ghazal Geshnizjani $^{2)}$\footnote{ghazal@physics.wisc.edu}}
\affiliation{1) Physics Department, Brown University,
  Providence RI 02912 USA,}
\affiliation{2) Department of Physics, University of Wisconsin, Madison WI 53706 USA.}
\date{\today}
\preprint{BROWN-HET-1449 }
\begin{abstract}
We point out an inconsistency in a method used in the literature
for studying adiabatic scalar perturbations in a regular bouncing
universe (in four dimensions). The method under scrutiny consists
of splitting the Bardeen potential into two pieces with
independent evolutions, in order to avoid a singular behavior at
the boundaries of the region where the null energy condition (NEC)
is violated. However, we argue that this method violates
energy-momentum conservation.

We then introduce a novel method
which provides two independent solutions for the Bardeen potential around the boundaries,
 even in the case of adiabatic perturbations. The two solutions are well behaved and not divergent.
\end{abstract}
\maketitle

In recent years, regular toy-models of a bouncing universe have received a lot of attention due
to string inspired singular bouncing models such as the cyclic scenario. Of particular interest
is the evolution of scalar perturbations, since their spectrum is directly observable in the
 anisotropies of the cosmic background radiation and the large scale structure of the universe.
 The majority of toy models are four dimensional, requiring at least two matter fields,
  of which at least one has to violate the null energy condition (NEC).

If one focuses on adiabatic perturbations, the evolution equation for the Bardeen potential becomes singular at
the boundaries of the NEC violating region before and after the bounce.
One way out of this dilemma was advocated in \cite{Peter:2002cn} and subsequently used e.g. in \cite{Peter:2003,Finelli:2003mc,Battefeld:2005cj}:
One splits the Bardeen potential in two components, each of which satisfies a regular second order
differential equation.

In this note we first show that this method is inconsistent because the fluid conservation equations are violated.
Therefore all models using this method have to be re-evaluated. We then introduce a novel method,
providing two independent and well behaved solutions for the Bardeen potential around the boundaries of the region with NEC violation,
 even in the case of adiabatic perturbations.

For simplicity, we work with a two fluid model with $
T^{\mu}_{\,\,\nu}= T^{\mu}_{(a)\nu}\pm T^{\mu}_{(b)\nu}$
and
equations of state $p_{(l)}=w_l\rho_{(l)}$ for $l=a,b$.
Note that a bounce for the scale factor $a$ occurs only in case of a negative sign in front of $T^{\mu}_{(b)\nu}$  \footnote{$w_b=2w_a+1$ corresponds
to the bounce in \cite{Battefeld:2005cj} and $w_b=2w_a+1/3$
corresponds to the one in \cite{Finelli:2003mc}.}.
Perturbing the metric
\begin{eqnarray}
d\,s^2&=&a^2\left[(1+2\Phi)d\,\eta ^2 -(1-2\Phi)\delta _{ij}d\,x^id\,x^j\right]\,,
\end{eqnarray}
where $\Phi$ is the Bardeen potential (longitudinal gauge,
 no anisotropic stress, see \cite{Mukhanov:1990me} for details),
and perturbing also the energy momentum tensors
\begin{eqnarray}
( \delta T^{\mu}_{(l)\nu})&=&\rho _{(l)}
\left(
\begin{array}{cc}
-\xi_{(l)}&(1+w_l)V_{(l),i}\\
-(1+w_l)V_{(l),i}&(w_l\xi_{(l)}) \delta ^i_j
\end{array}
\right)
\end{eqnarray}
where $\delta p_{(l)}=w_l\delta\rho_{(l)}$ and $\xi_{(l)} :=\delta \rho_{(l)}/ \rho _{(l)}$,
the perturbed Einstein equations for $\mathcal{K}=0$ (a spatially flat universe) read
\begin{eqnarray}
&&\nonumber \nabla ^2\Phi-3\mathcal{H}(\mathcal{H}\Phi +\Phi ^{\prime})=\frac{a^2}{2}\kappa^2\left(\rho _{(a)}\xi_{(a)}\pm\rho _{(b)}\xi_{(b)}\right)\,,\\
\label{Einstein1}\\
&&\nonumber \Phi ^{\prime\prime}+3\mathcal{H}\Phi ^{\prime}+(2\mathcal{H}^{\prime}+\mathcal{H}^2)\Phi=
\frac{a^2}{2}\kappa^2\big(w_a\xi_{(a)}\rho_{(a)}\\
&&\hspace{4.65cm}\pm w_b\xi_{(b)}\rho_{(b)}\big)\,,
\label{Einstein2}\\
\nonumber &&\left[\Phi\mathcal{H}+\Phi ^{\prime}\right]_{,i}=-\frac{a^2}{2}\kappa^2
\big(\rho_{(a)}V_{(a),i}(1+w_a)\\
&&\hspace{2.2cm}\pm
\rho_{(b)}V_{(b),i}(1+w_b)\big)\,,\label{Einstein3}
\end{eqnarray}
with $\kappa^{2}=8\pi /M_p^2$. The energy conservation equations
for each fluid, assuming no non-gravitational interactions between
the fluids, are
\begin{eqnarray}
\rho_{(l)}\left((1+w_l)\left[\nabla ^2 V_{(l)}-3\Phi
^{\prime}\right]+\xi_{(l)} ^\prime\right)=0\,. \label{cons}
\end{eqnarray}
In the long wavelength limit the above relation yields
\begin{eqnarray}
{\xi_{(l)} ^\prime \over 1+w_l}=3\Phi^{\prime}\,,
\label{entropycon}
\end{eqnarray}
which is another way of stating entropy conservation, that is
\begin{eqnarray}
S^\prime={\xi_{(a)} ^\prime \over 1+w_a}- {\xi_{(b)} ^\prime \over
1+w_b}=0\,.
\end{eqnarray}
One can simply impose the adiabaticity condition by setting
initially $S(\eta_{in}):=0$ and thus \footnote{While one can not
use the energy conservation equations for separate fluids in the
case discussed in \cite{Battefeld:2005cj} to derive this
condition, one can still arrive at the same relation due to the
specific form of the corrections to the energy momentum tensor in
\cite{Battefeld:2005cj}.}
\begin{eqnarray}
{\xi_{(a)} \over 1+w_a}={\xi_{(b)}\over 1+w_b} \label{adiabcon}
\end{eqnarray}
has to hold.

Substituting this result  back into (\ref{Einstein1}) and
(\ref{Einstein2}) and then combining the two, one arrives at a
second order equation for $\Phi$:
\begin{eqnarray}
&&0=\Phi^{\prime\prime} \bigg[\rho_a(1+w_a)\pm\rho_b(1+w_b)\bigg ]\label{eomPhi}\\
\nonumber&& +3\mathcal{H}\Phi^{\prime}\bigg[\rho_a(1+w_a)^2\pm \rho_b(1+w_b)^2\bigg ]\\
\nonumber&&+\bigg [-\big( w_a(w_a+1)\rho_a\pm w_b(w_b+1)\rho_b\big)\nabla^2\\
\nonumber &&+2\mathcal{H}^{\prime}\big(\rho_a(1+w_a)\pm\rho_b(1+w_b)\big)\\
&&\nonumber +\mathcal{H}^2 \big(\rho_a(1+w_a)(1+3w_a)
\pm\rho_b(1+w_b)(1+3w_b)\big) \bigg]\Phi\,.
\end{eqnarray}
In case of a minus sign, which is needed for a bounce to occur, or some negative $w_i$ this equation becomes singular at the boundaries of the NEC violating region
(see also \cite{Finelli:2003mc}).

In the appendix of \cite{Peter:2002cn} a split of $\Phi$ into $\Phi_a + \Phi_b$ is suggested, such that each $\Phi_{l}$ satisfies
\begin{eqnarray}
&&\nabla ^2\Phi_l-3\mathcal{H}(\mathcal{H}\Phi_l +\Phi_l
^{\prime})=\frac{a^2}{2}\kappa^2\rho _{(l)}\xi_{(l)},
\label{Einsteinphi1}\\
&&\Phi_l ^{\prime\prime}+3\mathcal{H}\Phi_l
^{\prime}+(2\mathcal{H}^{\prime}+\mathcal{H}^2)\Phi_l=
\frac{a^2}{2}\kappa^2w_l\xi_{(l)}\rho_{(l)} \label{Einsteinphi2}\,.
\end{eqnarray}
By adding these equations one arrives at the original Einstein
equations (\ref{Einstein1}) and (\ref{Einstein2}). Combining
(\ref{Einsteinphi1}) and (\ref{Einsteinphi2}) one can derive
for each $\Phi_l$ the equation
\begin{eqnarray}
\nonumber 0&=&\Phi^{\prime\prime}_l +3\mathcal{H}(1+w_l)\Phi^{\prime}_l\\
&&+\left(-w_l\nabla^2+2\mathcal{H}^{\prime}+(1+3w_l)\mathcal{H}^2\right)\Phi_l\label{eomPhil}\,.
\end{eqnarray}
Note that these equations are regular for each $\Phi_l$. This is
already the first hint that the method is doubtful, since the
singular behavior vanished miraculously. Another reason to doubt
this method is the fact that by just using equations
(\ref{Einstein1}) and (\ref{Einstein2}), with no extra constraint
like conservation of energy for each fluid or adiabaticity, one
seems to be able to calculate the evolution of each $\Phi_l$ and
subsequently $\Phi$ itself. However, it is obvious that in the
equations (\ref{Einstein1}) and (\ref{Einstein2}) three unknowns
appear, so that two equations are insufficient to calculate their
evolution. Therefore one can not expect the solutions of
(\ref{Einsteinphi1}) and (\ref{Einsteinphi2}) to be consistent
with the conservation of the energy-momentum tensor or other
constraint equations.

One easy way to see this inconsistency is to look
at the solutions in the long wavelength limit where equation
(\ref{Einsteinphi1}) simplifies to
\begin{eqnarray}
&&-3\mathcal{H}(\mathcal{H}\Phi_l +\Phi_l
^{\prime})=\frac{a^2}{2}\kappa^2\rho _{(l)}\xi_{(l)}\,.
\label{Einsteinphi1IR}
\end{eqnarray}
Adding the time derivative of the above equation to $\mathcal{H}\times$ itself and $3\mathcal{H}\times$ (\ref{Einsteinphi2})
yields
\begin{eqnarray}
\rho_{(a)}\xi_{(a)}
^\prime&=&3\Phi_{a}^\prime\left(\rho_{(a)}(1+w_a)+\rho_{(b)}(1+w_b)\right)\,,\label{fakecon2}\\
\rho_{(b)}\xi_{(b)}
^\prime&=&3\Phi_{b}^\prime\left(\rho_{(b)}(1+w_b)+\rho_{(a)}(1+w_a)\right)\,,
\label{fakecon2}
\end{eqnarray}
where we also used the background Einstein equations and $\rho^\prime_i=-3\mathcal{H}(1+w_i)\rho_i$.
This result, together with the conservation equation
(\ref{entropycon}), implies that
 the solutions of
(\ref{eomPhil}) have to satisfy
\begin{eqnarray}
{\Phi_a^\prime \over \rho_{(a)}(1+w_a)}={\Phi_b^\prime
\over\rho_{(b)}(1+w_b)}\,.\label{ccc}
\end{eqnarray}
One might think that this constraint is satisfied in the case of
adiabatic perturbations, but it is not as we shall see now. Using
(\ref{adiabcon}) in (\ref{Einsteinphi1IR}) yields
\begin{eqnarray}
\frac{\mathcal{H}\Phi_a+\Phi_a^\prime}{\rho_{(a)}(1+w_a)}=\frac{\mathcal{H}\Phi_b+\Phi_b^\prime}{\rho_{(b)}(1+w_b)}\,,
\end{eqnarray}
which can further be simplified by using (\ref{ccc}) to
\begin{eqnarray}
\frac{\Phi_a}{\rho_{(a)}(1+w_a)}=\frac{\Phi_b}{\rho_{(b)}(1+w_b)}\,.\label{cccc}
\end{eqnarray}
This together with (\ref{ccc}) implies
\begin{eqnarray}
\frac{\Phi_a^\prime}{\Phi_a}=\frac{\Phi_b^\prime}{\Phi_b}\,,
\end{eqnarray}
yielding $\Phi_a\propto\Phi_b$. However, this is clearly in contradiction to (\ref{cccc}), since the densities have a different dependency on conformal time if $w_a\neq w_b$.
Therefore the splitting method itself is inconsistent and can not
be trusted to regularize (\ref{eomPhi}).

We propose another mathematical technique that can be used instead
to approximate the solutions of (\ref{eomPhi}) in the vicinity of
$\rho_{tot}+p_{tot}=0$ at $\eta_{nec}$. Equation (\ref{eomPhi}) is
a second order differential equation that in fourier space
(suppressing the subscript on $\Phi_k$), has the following general
form:
\begin{eqnarray}
\label{geneq}A(\eta)\Phi^{\prime\prime}+B(\eta)\Phi^\prime+C(k,\eta)\Phi=0,
\end{eqnarray}
where $A$ and $B$ are related via
\begin{eqnarray}
\label{AB}
B=-A^{\prime}\,,
\end{eqnarray}
since energy conservation for each
fluid requires \footnote{Again, although the energy conservation
argument does not hold for \cite{Battefeld:2005cj} this
relation still remains valid.}
\begin{eqnarray}
\label{encon}
\rho^\prime_l=-3\mathcal{H}\rho_{_{l}}(1+w_l).\end{eqnarray}

Our first goal is to derive one of the solutions around
$\eta_{nec}$ perturbatively. We can then obtain the second solution by means of the Wronskian method.

 Since $a(\eta_{nec})\neq 0$ we can Taylor expand $\rho_l$
 and consequently $A, B$ and $C$ around $\eta_{nec}$:
 \begin{eqnarray}
&& A(\delta)=A_1\delta+A_2\delta^2+A_3\delta^3+\ldots\,, \label{A}\\
 &&B(\delta)=-A_1-2A_2\delta-3A_3\delta^2+\ldots\,,\\
 &&C(\delta)=C_0+C_1\delta+C_2\delta^2+\ldots\,,
\end{eqnarray}
where we defined $\delta=\eta-\eta_{nec}$ and used
$A(\eta_{nec})=0$. We assume that $A_1$, corresponding to the
linear term of $A(\delta)$, does not vanish. This is usually the
case and plays a crucial role in our analysis.

Furthermore, we assume the existence of an analytic solution around $\eta_{nec}$ (we will see bellow that it is a
justified assumption \footnote{Arguments in favor of an analytic
$\Phi$ where given in e.g. \cite{Peter:2003} or
\cite{Bozza:2005xs}, without relying on the splitting method. }),
so that it can be written as
\begin{eqnarray}\label{sol1}
\Phi_1=\alpha_0+\alpha_1\delta+\alpha_2\delta^2+\ldots\,.
\end{eqnarray}
Substituting relations (\ref{A})-(\ref{sol1}) into (\ref{geneq})
we obtain the following relations from the zeroth and first order
equations in $\delta$
\begin{eqnarray}
\mathcal{O}(\delta^0) && \Rightarrow ~~~~ 0=-A_1\alpha_1+C_0\alpha_0 \,,
\label{O0}\\
\mathcal{O}(\delta^1) &&\Rightarrow ~~~~ 0=(-2A_2+C_0)\alpha_1+C_1\alpha_0
\,. \label{O1}
\end{eqnarray}
In general, this can only be satisfied if
$\alpha_0=\alpha_1=0$. Fortunately, this does not imply
$\Phi_1\equiv 0$, because the
 equations of higher order in $\delta$ can all be
satisfied recursively. In fact, we can compute a complete power
series solution for $\Phi_1$
\begin{eqnarray}
\mathcal{O}(\delta^2)  \!&\Rightarrow&\!  \alpha_3\!=\!{2A_2-C_0\over 3A_1}~ \alpha_2
\label{O2}\\
&\vdots&  \nonumber\\
\mathcal{O}(\delta^n)  \!&\Rightarrow&\! \alpha_{n+1}\!=\!{\sum_{i=2}^n[i(3+n-2i)A_{n+2-i}-C_{n-i}]\alpha_i\over(n-1)(n+1)A_1} \nonumber
\end{eqnarray}
with the consequence
\begin{eqnarray}\label{Phi1}
\Phi_1= \delta^2+{2A_2-C_0\over 3A_1}\delta^3+\ldots\,.
\end{eqnarray}
Knowing one of the solutions of (\ref{geneq}), $\Phi_1$, we can easily obtain the other solution, $\Phi_2$, by using the Wronskian
technique. The Wronskian for a second order differential equations is defined as
\begin{eqnarray}
W=\Phi^\prime_2\Phi_1-\Phi^\prime_1\Phi_2 \label{w}\,,
\end{eqnarray}
where $\Phi_1$ and $\Phi_2$ are the independent solutions of
(\ref{geneq}). Henceforth, one can calculate $\Phi_2$ in
terms of $W$ and $\Phi_1$:
\begin{eqnarray} \label{phi2w}
\Phi_2\sim\Phi_1\int {W\over \Phi_1^2}d\eta\,.
\end{eqnarray}
$W$ itself satisfies the first order differential equation
\begin{eqnarray}\label{weq}
A(\eta)W^\prime+B(\eta)W=0\,.
\end{eqnarray}
By substituting $B(\eta)$ from (\ref{AB}) we can solve the above equation for $W$ to
\begin{eqnarray}
W(\eta)=\beta A(\eta),
\end{eqnarray}where $\beta$ is just a constant. Combining this result for $W(\eta)$ with (\ref{phi2w}) and
our solution for $\Phi_1$ from (\ref{Phi1}), we end up with
\begin{eqnarray}\label{phi2}
\Phi_2=-{A_1\over2}-{8A_2-C_0\over6}\delta+A_3\delta^2\ln(|\delta|)+O(\delta^2).
\end{eqnarray}
Note that although this solution is not analytic at $\eta=\eta_{nec}$  ($\delta=0$),
it is well behaved in the sense that both, the solution and its first derivative, remain continuous and finite.
Thus, we have constructed the approximate form of the two independent solutions of (\ref{geneq})
or subsequently (\ref{eomPhi}). These can be used to match the solution on different sides of $\eta_{nec}$.

To summarize, we have shown explicitly that the splitting method, first introduced in \cite{Peter:2002cn} and used to show the regularity of the Bardeen potential, is intrinsically inconsistent.
All models using this method have to be re-evaluated \footnote{The model in \cite{Battefeld:2005cj}
V.2 by the authors of this critique will be corrected shortly in the upcoming revision V.3.
The background solution in \cite{Battefeld:2005cj}, as well as the sections on vector and tensor
perturbations are unaffected by this revision.}, e.g. by using the technique introduced in this draft or by
including entropy perturbations, as emphasized in \cite{Peter:2003} and
later on in \cite{PintoNeto:2004wf} (they derived a regular forth order equation for the full
Bardeen potential).
\begin{acknowledgments}
We would like to thank the referee of \cite{Battefeld:2005cj} for drawing our attention to a possible problem associated with the splitting method, and N.~Afshordi, D.~Battefeld,  R.~Brandenberger and D.~Chung for comments
on the draft.
\end{acknowledgments}

\end{document}